\documentclass[12pt,hyperref,tightenlines,showpacs]{revtex4}%
\usepackage{amsfonts}
\usepackage[debug,pdftex]{hyperref}
\usepackage{amsmath}
\usepackage{amssymb}
\usepackage{graphicx}
\usepackage{bibmods}%
\providecommand{\U}[1]{\protect\rule{.1in}{.1in}}

\begin{document}
\title{Path integral solution for an angle-dependent anharmonic oscillator}
\author{S. Haouat}
\email{s\_haouat@univ-jijel.dz}
\affiliation{\textit{LPTh, Department of Physics, University of Jijel, BP 98, Ouled Aissa,
18000 Jijel, Algeria.}}

\begin{abstract}
We have given a straightforward method to solve the problem of noncentral
anharmonic oscillator in three dimensions. The relative propagator is
presented by means of path integrals in spherical coordinates. By making an
adequate change of time we were able to separate the angular motion from the
radial one. The relative propagator is then exactly calculated. The energy
spectrum and the corresponding wave functions are obtained.

\end{abstract}

\maketitle

\section{Introduction\qquad}

As we know the Feynman path integral formalism has played an important role in
the comprehension of the quantum phenomena \cite{P1,P2,P3}. In nonrelativistic
quantum mechanics, many problems have been exactly solved starting from their
classical origins by the use of path integrals \cite{P4,P5}.\ Therefore, it
has been proved that this method is powerful in finding exact propagators and
wave functions and studying quantum behavior of several systems \cite{P6}. At
a certain epoch, however, the development of the path integral method wasn't
rapid compared with Schr\"{o}dinger and Heisenberg approaches. The
introduction of the space and time transformation to solve the nonrelativistic
Coulomb problem and other basic potentials was the decisive stage for its
development \cite{Duru,P7}. Having crossed this crucial stage, the path
integral quantization method becomes at the present time more important than
the Schr\"{o}dinger equation. Its rise unveils itself from its applications in
several fields of physics such as quantum mechanics \cite{P1,P2,P3,P4,P5,P6},
statistical physics \cite{P7}, quantum field theory \cite{P8,P9}, condensed
matter \cite{P10}, cosmology \cite{P11,P12}and black hole physics \cite{P13}.
We can find also a modern introduction to path integrals with significant
applications in \cite{Chaichian1,Chaichian2}. This explains the increasing
interest in the development of path integration techniques.

However to our knowledge there is no report on the path integral treatment for
a class of noncentral potentials which are of interest in understanding the
nuclear shell structure within the framework of a modified mean field model
\cite{setare}. These potentials that describe the rotational-vibrational
motion of the nuclear system, have the form%
\begin{equation}
V\left(  r,\theta\right)  =\frac{1}{2}\mu\omega^{2}r^{2}+q~\frac{u\left(
\theta\right)  }{r^{2}} \label{1}%
\end{equation}
where $\frac{1}{2}\mu\omega^{2}r^{2}$ is the usual three dimensional harmonic
oscillator that explains the magic numbers for spherical symmetric nucleus,
$u\left(  \theta\right)  $ is a function of $\theta$ and $q$ is a deformation
parameter. We note that a particular case in two dimensional space is
discussed in \cite{Grosch}.

In the present paper, we use of the Feynman path integral method to solve the
problem of a nonrelativistic particle subjected to the following noncentral
potential%
\begin{equation}
V\left(  r,\theta\right)  =-V_{0}+\frac{1}{2}\mu\omega^{2}r^{2}+\frac
{\alpha\hbar^{2}}{2\mu r^{2}}+\frac{\beta\hbar^{2}\cos^{2}\theta}{2\mu
r^{2}\sin^{2}\theta}+\frac{\gamma\hbar^{2}}{2\mu r^{2}\cos^{2}\theta},
\label{2}%
\end{equation}
where $V_{0}$, $\alpha$, $\beta$ and $\gamma$ are constants. We note that some
particular cases of this potential are studied in \cite{NC1,NC2,NC3}. Also
some similar potentials have been recently discussed either in nonrelativistic
quantum mechanics or in relativistic theory \cite{NC4,NC5,NC6,NC7,NC8}.

In the first stage we formulate the problem in spherical coordinates path
integrals. Then by the use of a temporal transformation we separate the
angular motion and we do integration over $\theta$ and $\rho$ to obtain the
exact propagator. Finally, we extract the energy spectrum and the wave functions.

\section{Path integral formulation}

In nonrelativistic quantum mechanics, the propagator or the transition
amplitude from initial state $\left\vert \vec{r}_{a}\right\rangle $ to final
state $\left\vert \vec{r}_{b}\right\rangle $ for a physical system governed by
the Hamiltonian $H$ and the Lagrangian $\mathcal{L}$, is defined, in
configuration space, by a matrix element of the evolution operator%

\begin{equation}
K\left(  \vec{r}_{b},t_{b};\vec{r}_{a},t_{a}\right)  =\left\langle \vec{r}%
_{b}\left\vert \exp\left[  -\frac{i}{\hbar}H\left(  t_{b}-t_{a}\right)
\right]  \right\vert \vec{r}_{a}\right\rangle , \label{3}%
\end{equation}
which admits the following functional integral representation%

\begin{equation}
K\left(  \vec{r}_{b},t_{b};\vec{r}_{a},t_{a}\right)  =%
{\displaystyle\int}
D\vec{r}(t)\exp\left(  \frac{i}{\hbar}S\left[  \vec{r}(t)\right]  \right)  ,
\label{4}%
\end{equation}
where the action $S\left[  \vec{r}(t)\right]  $ is a functional of continuous
trajectory $\vec{r}(t)$ connecting space-time point $\left(  \vec{r}_{a}%
,t_{a}\right)  $ with $\left(  \vec{r}_{b},t_{b}\right)  $%

\begin{equation}
S\left[  \vec{r}(t)\right]  =\int_{t_{a}}^{t_{b}}dtL(\vec{r},\overset{\cdot
}{\vec{r}}),\label{5}%
\end{equation}
with the Lagrangian
\begin{equation}
L=\frac{\mu}{2}\left(  \dot{x}^{2}+\dot{y}^{2}+\dot{z}\right)  -\tilde
{V}\left(  x,y,z\right)  \label{6}%
\end{equation}
and the potential%
\begin{equation}
\tilde{V}\left(  x,y,z\right)  =-V_{0}+\frac{1}{2}\mu\omega^{2}\left(
x^{2}+y^{2}+z^{2}\right)  +\frac{\left(  \alpha-\beta\right)  \hbar^{2}}%
{2\mu\left(  x^{2}+y^{2}+z^{2}\right)  }+\frac{\beta\hbar^{2}}{2\mu\left(
x^{2}+y^{2}\right)  }+\frac{\gamma\hbar^{2}}{2\mu z^{2}}\label{7}%
\end{equation}
In its discrete form $K\left(  \vec{r}_{b},t_{b};\vec{r}_{a},t_{a}\right)  $
can be written as
\begin{equation}
K\left(  \vec{r}_{b},t_{b};\vec{r}_{a},t_{a}\right)  \equiv K\left(  \vec
{r}_{b},\vec{r}_{a};T\right)  =\int Dx\int Dy\int Dz\exp\left(  \frac{i}%
{\hbar}\sum_{n}A_{n}\right)  ,\label{8}%
\end{equation}
where%
\begin{equation}
A_{n}=\left(  \frac{\mu}{2}\frac{\left(  \Delta x_{n}\right)  ^{2}+\left(
\Delta y_{n}\right)  ^{2}+\left(  \Delta z_{n}\right)  ^{2}}{\varepsilon
}-\varepsilon\tilde{V}\left(  x_{n},y_{n},z_{n}\right)  \right)  \label{9}%
\end{equation}
and
\begin{equation}
\int Dx\int Dy\int Dz=\lim_{\substack{N\rightarrow\infty\\\varepsilon
\rightarrow0}}\int\prod\limits_{n=1}^{N-1}dx_{n}\int\prod\limits_{n=1}%
^{N-1}dy_{n}\int\prod\limits_{n=1}^{N-1}dz_{n}~\int\prod\limits_{n=1}%
^{N}\left(  \sqrt{\frac{\mu}{2i\pi\hbar\varepsilon}}\right)  ^{3}.\label{10}%
\end{equation}
Here we have used the standard notation%
\begin{equation}%
\begin{array}
[c]{l}%
\Delta q_{n}=q_{n}-q_{n-1}\\
q_{N}=q\left(  t_{b}\right)  \\
q_{0}=q\left(  t_{a}\right)  \\
T=t_{b}-t_{a}=N\varepsilon.
\end{array}
\label{11}%
\end{equation}
Let us now search for a path integral representation in the spherical
coordinates%
\begin{equation}%
\begin{array}
[c]{l}%
x_{n}=r_{n}\sin\theta_{n}\cos\varphi_{n}\\
y_{n}=r_{n}\sin\theta_{n}\sin\varphi_{n}\\
z_{n}=r_{n}\cos\theta_{n}.
\end{array}
\label{12}%
\end{equation}
It is obvious that the measure%
\[
\int\prod\limits_{n=1}^{N-1}dx_{n}\int\prod\limits_{n=1}^{N-1}dy_{n}\int
\prod\limits_{n=1}^{N-1}dz_{n}~\prod\limits_{n=1}^{N}\left(  \sqrt{\frac{\mu
}{2i\pi\hbar\varepsilon}}\right)  ^{3}%
\]
will be in spherical coordinates
\[
\int\prod\limits_{n=1}^{N-1}dr_{n}\int\prod\limits_{n=1}^{N-1}d\theta_{n}%
\int\prod\limits_{n=1}^{N-1}d\varphi_{n}\prod\limits_{n=1}^{N-1}r_{n}^{2}%
\sin\theta_{n}\sqrt{\left(  \frac{\mu}{2i\pi\hbar\varepsilon}\right)  ^{3N}}~
\]
and the propagator takes the following form%

\begin{align}
K\left(  \vec{r}_{b},\vec{r}_{a};T\right)   &  =\lim_{\substack{N\rightarrow
\infty\\\varepsilon\rightarrow0}}\int\prod\limits_{n=1}^{N-1}dr_{n}\int
\prod\limits_{n=1}^{N-1}d\theta_{n}\int\prod\limits_{n=1}^{N-1}d\varphi
_{n}\prod\limits_{n=1}^{N-1}r_{n}^{2}\sin\theta_{n}\sqrt{\left(  \frac{\mu
}{2i\pi\hbar\varepsilon}\right)  ^{3N}}\nonumber\\
&  \exp\sum_{n}\left(  \frac{i}{\hbar}\frac{\mu}{2}\frac{r_{n}^{2}+r_{n-1}%
^{2}-2r_{n}r_{n-1}\cos\theta_{n}\cos\theta_{n-1}}{\varepsilon}\right.
\nonumber\\
&  \left.  -\frac{i}{\hbar}\frac{\mu r_{n}r_{n-1}\sin\theta_{n}\sin
\theta_{n-1}}{\varepsilon}\cos\left(  \Delta\varphi_{n}\right)  -\frac
{i}{\hbar}\varepsilon V\left(  r_{n},\theta_{n}\right)  \right)  .\label{13}%
\end{align}
Taking into account that%
\begin{equation}
\prod\limits_{n=1}^{N-1}r_{n}^{2}=\frac{1}{r_{b}r_{a}}\prod\limits_{n=1}%
^{N}\left(  r_{n}r_{n-1}\right)  \label{14}%
\end{equation}
and using the following formulae%
\begin{equation}
\exp\left(  -i\frac{a}{\varepsilon}\cos\left(  \Delta\varphi_{n}\right)
\right)  =\sum_{-\infty}^{+\infty}I_{m}\left(  -i\frac{a}{\varepsilon}\right)
e^{im\Delta\varphi_{n}},\label{15}%
\end{equation}
where $I_{m}\left(  x\right)  $\ is the modified Bessel function, we obtain%
\begin{align}
K\left(  \vec{r}_{b},\vec{r}_{a};T\right)   &  =\frac{1}{r_{b}r_{a}\sqrt
{\sin\theta_{b}\sin\theta_{a}}}\nonumber\\
&  \int\prod\limits_{n=1}^{N-1}dr_{n}\int\prod\limits_{n=1}^{N-1}d\theta
_{n}\int\prod\limits_{n=1}^{N-1}d\varphi_{n}\prod\limits_{n=1}^{N}\left(
r_{n}r_{n-1}\sqrt{\sin\theta_{n}\sin\theta_{n-1}}\right)  \sqrt{\left(
\frac{\mu}{2i\pi\hbar\varepsilon}\right)  ^{3N}}\nonumber\\
&  \prod\limits_{n=1}^{N}\sum_{m=-\infty}^{+\infty}I_{m}\left(  -i\mu
\frac{r_{n}r_{n-1}}{\hbar\varepsilon}\sin\theta_{n}\sin\theta_{n-1}\right)
e^{im\Delta\varphi_{n}}\nonumber\\
&  \exp\left(  \sum_{n}\left(  \frac{i}{\hbar}\frac{\mu}{2}\frac{r_{n}%
^{2}+r_{n-1}^{2}-2r_{n}r_{n-1}\cos\theta_{n}\cos\theta_{n-1}}{\varepsilon
}-\frac{i}{\hbar}V\left(  r_{n},\theta_{n}\right)  \right)  \varepsilon
\right)  .\label{16}%
\end{align}
For small $\varepsilon,$ the modified Bessel function behaves as follows%

\begin{equation}
I_{m}\left(  -i\frac{a}{\varepsilon}\right)  \approx\left(  \frac
{i\varepsilon}{2\pi a}\right)  ^{\frac{1}{2}}\times\exp\left[  -i\frac
{a}{\varepsilon}-\frac{i\varepsilon}{2a}\left(  m^{2}-\frac{1}{4}\right)
\right]  ,\label{17}%
\end{equation}
what permits to us to write%
\begin{align}
K\left(  \vec{r}_{b},\vec{r}_{a};T\right)   &  =\frac{1}{2\pi r_{b}r_{a}%
\sqrt{\sin\theta_{b}\sin\theta_{a}}}\nonumber\\
&  \int\prod\limits_{n=1}^{N-1}dr_{n}\int\prod\limits_{n=1}^{N-1}d\theta
_{n}\prod\limits_{n=1}^{N}\left(  r_{n}r_{n-1}\right)  ^{\frac{1}{2}}%
\sqrt{\left(  \frac{\mu}{2i\pi\hbar\varepsilon}\right)  ^{2N}}\int
\prod\limits_{n=1}^{N-1}\frac{d\varphi_{n}}{2\pi}\nonumber\\
&  \prod\limits_{n=1}^{N}\sum_{m=-\infty}^{+\infty}\exp\left[  im\Delta
\varphi_{n}+\frac{i}{\hbar}\frac{\mu}{2}\frac{r_{n}^{2}+r_{n-1}^{2}%
-2r_{n}r_{n-1}\cos\left(  \Delta\theta_{n}\right)  }{\varepsilon}\right.
\nonumber\\
&  \left.  -i\frac{\left(  m^{2}-\frac{1}{4}\right)  \hbar\varepsilon}{2\mu
r_{n}r_{n-1}\sin\theta_{n}\sin\theta_{n-1}}-\frac{i}{\hbar}V\left(
r_{n},\theta_{n}\right)  \varepsilon\right]  .\label{18}%
\end{align}
Now by using the Taylor development up to order 4
\begin{equation}
\cos\Delta\theta_{n}=\allowbreak1-\frac{1}{2}\left(  \Delta\theta_{n}\right)
^{2}+\frac{1}{24}\left(  \Delta\theta_{n}\right)  ^{4}+...\label{19}%
\end{equation}
we have%
\begin{align}
K\left(  \vec{r}_{b},\vec{r}_{a};T\right)   &  =\frac{1}{2\pi r_{b}r_{a}%
\sqrt{\sin\theta_{b}\sin\theta_{a}}}\\
&  \int\prod\limits_{n=1}^{N-1}dr_{n}\int\prod\limits_{n=1}^{N-1}d\theta
_{n}\prod\limits_{n=1}^{N}\left(  \frac{\mu r_{n}r_{n-1}}{2i\pi\hbar
\varepsilon}\right)  ^{\frac{1}{2}}\sqrt{\left(  \frac{\mu}{2i\pi
\hbar\varepsilon}\right)  ^{N}}\nonumber\\
&  \int\prod\limits_{n=1}^{N-1}\frac{d\varphi_{n}}{2\pi}\prod\limits_{n=1}%
^{N}\sum_{m=-\infty}^{+\infty}\exp\frac{i}{\hbar}\left[  m\hbar\Delta
\varphi_{n}+\frac{\mu}{2}\frac{\left(  \Delta r_{n}\right)  ^{2}}{\varepsilon
}+\frac{\mu}{2}r_{n}r_{n-1}\frac{\left(  \Delta\theta_{n}\right)  ^{2}%
}{\varepsilon}\right.  \nonumber\\
&  \left.  -\frac{\mu}{24}\frac{r_{n}r_{n-1}}{\varepsilon}\left(  \Delta
\theta_{n}\right)  ^{4}-\frac{\left(  m^{2}-\frac{1}{4}\right)  \hbar
^{2}\varepsilon}{2\mu r_{n}r_{n-1}\sin\theta_{n}\sin\theta_{n-1}}-V\left(
r_{n},\theta_{n}\right)  \varepsilon\right]  .\label{20}%
\end{align}
According to McLaughlin-Shulman procedure \cite{mclaughlin}, the term with
$\left(  \Delta\theta_{n}\right)  ^{4}$ leads to a quantum correction that can
be calculated with the help of the property%
\begin{equation}
\int\prod\limits_{n=1}^{N-1}d\theta_{n}=\int\prod\limits_{n=2}^{N}d\left(
\Delta\theta_{n}\right)  \label{21}%
\end{equation}
and the use of the integral%
\begin{equation}
\int du~u^{4}\exp\left[  iau^{2}\right]  =\left(  \frac{-3}{4a^{2}}\right)
\int du~\exp\left[  iau^{2}\right]  .\label{22}%
\end{equation}
This correction is%
\begin{equation}
\left\langle \left(  \Delta\theta_{n}\right)  ^{4}\right\rangle \approx
3\left(  \frac{i\hbar\varepsilon}{\mu r_{n}r_{n-1}}\right)  ^{2}.\label{23}%
\end{equation}
Then the propagator $K\left(  \vec{r}_{b},\vec{r}_{a};T\right)  $ takes the
following path integral representation
\begin{align}
K\left(  \vec{r}_{b},\vec{r}_{a};T\right)   &  =\frac{1}{2\pi r_{b}r_{a}%
\sqrt{\sin\theta_{b}\sin\theta_{a}}}\nonumber\\
&  \int\prod\limits_{n=1}^{N-1}dr_{n}\int\prod\limits_{n=1}^{N-1}d\theta
_{n}\prod\limits_{n=1}^{N}\left(  \frac{\mu r_{n}r_{n-1}}{2i\pi\hbar
\varepsilon}\right)  ^{\frac{1}{2}}\sqrt{\left(  \frac{\mu}{2i\pi
\hbar\varepsilon}\right)  ^{N}}\nonumber\\
&  \int\prod\limits_{n=1}^{N-1}\frac{d\varphi_{n}}{2\pi}\prod\limits_{n=1}%
^{N}\sum_{m=-\infty}^{+\infty}\exp\frac{i}{\hbar}\left[  m\hbar\Delta
\varphi_{n}+\frac{\mu}{2}\frac{\left(  \Delta r_{n}\right)  ^{2}}{\varepsilon
}+\frac{\mu}{2}r_{n}r_{n-1}\frac{\left(  \Delta\theta_{n}\right)  ^{2}%
}{\varepsilon}\right.  \nonumber\\
&  \left.  +\left(  \frac{\hbar^{2}}{8\mu r_{n}r_{n-1}}\right)  \varepsilon
-\frac{\left(  m^{2}-\frac{1}{4}\right)  \hbar^{2}\varepsilon}{2\mu
r_{n}r_{n-1}\sin\theta_{n}\sin\theta_{n-1}}-V\left(  r_{n},\theta_{n}\right)
\varepsilon\right]  .\label{24}%
\end{align}
At this level, we see that only the integration over $\varphi$ is
straightforward. One cannot do integration over $\theta$ and $r$ because of
the position dependent kinetic term $\frac{\mu}{2}r_{n}r_{n-1}\left(
\Delta\theta_{n}\right)  ^{2}$ and the term $\hbar^{2}\left(  2\mu
r_{n}r_{n-1}\sin\theta_{n}\sin\theta_{n-1}\right)  ^{-1}$. Let us in that case
do integration over $\varphi$. We obtain%

\begin{equation}
K\left(  \vec{r}_{b},\vec{r}_{a};T\right)  =\sum_{m=-\infty}^{+\infty}%
\frac{e^{im\left(  \varphi_{b}-\varphi_{a}\right)  }}{2\pi}K_{m}\left(
r_{b},\theta_{b};r_{a},\theta_{a};T\right)  \label{25}%
\end{equation}
where the novel propagator $K_{m}\left(  r_{b},\theta_{b};r_{a},\theta
_{a};T\right)  $ is given by
\begin{align}
&  \left.  K_{m}\left(  r_{b},\theta_{b};r_{a},\theta_{a};T\right)  =\frac
{1}{r_{b}r_{a}\sqrt{\sin\theta_{b}\sin\theta_{a}}}\right. \nonumber\\
&  \int\prod\limits_{n=1}^{N-1}dr_{n}\int\prod\limits_{n=1}^{N-1}d\theta
_{n}\prod\limits_{n=1}^{N}\left(  \frac{\mu r_{n}r_{n-1}}{2i\pi\hbar
\varepsilon}\right)  ^{\frac{1}{2}}\sqrt{\left(  \frac{\mu}{2i\pi
\hbar\varepsilon}\right)  ^{N}}\nonumber\\
&  \exp\frac{i}{\hbar}\sum_{n}\left(  \frac{\mu}{2}\frac{\left(  \Delta
r_{n}\right)  ^{2}}{\varepsilon}+\frac{\mu}{2}r_{n}r_{n-1}\frac{\left(
\Delta\theta_{n}\right)  ^{2}}{\varepsilon}+\frac{\hbar^{2}}{8\mu r_{n}%
r_{n-1}}\varepsilon\right. \nonumber\\
&  \left.  -\frac{\left(  m^{2}-\frac{1}{4}\right)  \hbar^{2}\varepsilon}{2\mu
r_{n}r_{n-1}\sin\theta_{n}\sin\theta_{n-1}}-V\left(  r_{n},\theta_{n}\right)
\varepsilon\right)  . \label{26}%
\end{align}
Here we note that the terms $\frac{\hbar^{2}}{8\mu r_{n}r_{n-1}}$ and
$\frac{\hbar^{2}}{2\mu r_{n}r_{n-1}\sin\theta_{n}\sin\theta_{n-1}}$ represent
an effective potential that describes the quantum corrections resulting from
the passage to spherical coordinates. In the next step we separate the angle
dependence from the radial one by using a simple time transformation.

\section{Separation of variables}

Having formulated the problem of angle-dependant anharmonic oscillator in the
framework of path integrals in spherical coordinates, let us proceed to find
exact solutions by doing separation of variables. First we define the fixed
energy amplitude $\tilde{K}_{m}\left(  r_{b},\theta_{b};r_{a},\theta
_{a};E\right)  $ to be the Fourier transform of $K_{m}\left(  r_{b},\theta
_{b};r_{a},\theta_{a};T\right)  $%

\begin{equation}
K_{m}\left(  r_{b},\theta_{b};r_{a},\theta_{a};T\right)  =\int_{-\infty
}^{+\infty}\frac{dE}{2\pi}e^{-i\frac{E}{\hbar}T}\tilde{K}_{m}\left(
r_{b},\theta_{b};r_{a},\theta_{a};E\right)  \label{27}%
\end{equation}
with%
\begin{align}
&  \left.  \tilde{K}_{m}\left(  r_{b},\theta_{b};r_{a},\theta_{a};E\right)
=\frac{1}{r_{b}r_{a}\sqrt{\sin\theta_{b}\sin\theta_{a}}}\right. \nonumber\\
&  \int_{-\infty}^{+\infty}dTe^{i\frac{E}{\hbar}T}\int\prod\limits_{n=1}%
^{N-1}dr_{n}\int\prod\limits_{n=1}^{N-1}d\theta_{n}\prod\limits_{n=1}%
^{N}\left(  \frac{\mu\sqrt{r_{n}r_{n-1}}}{2i\pi\hbar\varepsilon}\right)
\nonumber\\
&  \exp\left[  \frac{i}{\hbar}\sum_{n}\left(  \frac{\mu}{2}\frac{\left(
\Delta r_{n}\right)  ^{2}}{\varepsilon}+\frac{\mu}{2}r_{n}r_{n-1}\frac{\left(
\Delta\theta_{n}\right)  ^{2}}{\varepsilon}+\left(  \frac{\hbar^{2}}{8\mu
r_{n}r_{n-1}}\right)  \varepsilon\right.  \right. \nonumber\\
&  \left.  \left.  -\frac{\left(  m^{2}-\frac{1}{4}\right)  \hbar\varepsilon
}{2\mu r_{n}r_{n-1}\sin\theta_{n}\sin\theta_{n-1}}-V\left(  r_{n},\theta
_{n}\right)  \varepsilon\right)  \right]  . \label{28}%
\end{align}
In order to be able to separate angular motion from the radial one we use the
method of \cite{chetouani}. To begin we change the evolution time from $t$ to
$s$, with%
\begin{equation}
ds=\frac{dt}{r^{2}}. \label{29}%
\end{equation}
This transformation is equivalent to
\begin{equation}
\varepsilon=\sigma_{n}r_{n}r_{n-1} \label{30}%
\end{equation}
with a time interval%
\begin{equation}
S=\int_{0}^{T}\frac{dt}{r^{2}}. \label{31}%
\end{equation}
To incorporate this changes in the path integral representation of $\tilde
{K}_{m}\left(  r_{b},\theta_{b};r_{a},\theta_{a};E\right)  ,$ we start form
the identity
\begin{equation}
\int dS\delta\left(  S-\int_{0}^{T}\frac{dt}{r^{2}}\right)  =1 \label{32}%
\end{equation}
and by the use of
\begin{equation}
\delta\left(  y-f\left(  x\right)  \right)  =\frac{1}{f^{\prime}\left(
x\right)  }\delta\left(  f^{-1}\left(  y\right)  -x\right)  \label{33}%
\end{equation}
we get%
\begin{equation}
\int dSr^{2}\left(  T\right)  \delta\left(  T-\int_{0}^{S}r^{2}ds\right)  =1.
\label{34}%
\end{equation}
By inserting the later identity in (\ref{28}) and being aware of $r\left(
T\right)  =r_{b},$ we can write $\tilde{K}_{m}\left(  r_{b},\theta_{b}%
;r_{a},\theta_{a};E\right)  $ as an integral of two independent kernels
\begin{equation}
\tilde{K}_{m}\left(  r_{b},\theta_{b};r_{a},\theta_{a};E\right)  =\int
_{0}^{+\infty}dS~P_{E}\left(  r_{b},r_{a};S\right)  ~Q_{m}\left(  \theta
_{b},\theta_{a};S\right)  \label{36}%
\end{equation}
where%
\begin{align}
&  \left.  P_{E}\left(  r_{b},r_{a};S\right)  =\right. \nonumber\\
&  \frac{r_{b}}{r_{a}}\int\prod\limits_{n=1}^{N-1}dr_{n}\prod\limits_{n=1}%
^{N}\left(  \sqrt{\frac{\mu}{2i\pi\hbar\sigma_{n}r_{n}r_{n-1}}}\right)
\exp\left\{  \frac{i}{\hbar}\sum_{n}\left[  \frac{\mu}{2}\frac{\left(  \Delta
r_{n}\right)  ^{2}}{\sigma_{n}r_{n}r_{n-1}}\right.  \right. \nonumber\\
&  \left.  \left.  -\left(  \frac{1}{2}m\omega^{2}r_{n}^{2}+\frac{\left(
\alpha-\beta\right)  \hbar^{2}}{2\mu r_{n}^{2}}\right)  r_{n}r_{n-1}\sigma
_{n}+\left(  E+V_{0}\right)  r_{n}r_{n-1}\sigma_{n}-\left(  \frac{\hbar^{2}%
}{8\mu}\right)  \sigma_{n}\right]  \right\}  \label{37}%
\end{align}
and%
\begin{align}
Q_{m}\left(  \theta_{b},\theta_{a};S\right)   &  =\frac{1}{\sqrt{\sin
\theta_{b}\sin\theta_{a}}}\int\prod\limits_{n=1}^{N-1}d\theta_{n}%
\prod\limits_{n=1}^{N}\left(  \sqrt{\frac{\mu}{2i\pi\hbar\sigma_{n}}}\right)
\nonumber\\
&  \exp\left\{  \frac{i}{\hbar}\sum_{n}\left(  \frac{\mu}{2}\frac{\left(
\Delta\theta_{n}\right)  ^{2}}{\sigma_{n}}-\frac{\hbar^{2}}{2\mu}\left(
\frac{\lambda^{2}-\frac{1}{4}}{\sin^{2}\theta_{n}}+\frac{k^{2}-\frac{1}{4}%
}{\cos^{2}\theta_{n}}\right)  \sigma_{n}\right)  \right\}  . \label{38}%
\end{align}
with%
\begin{align}
\lambda &  =\sqrt{\beta+m^{2}}\label{39}\\
k  &  =\sqrt{\gamma+\frac{1}{4}} \label{40}%
\end{align}
In the continuous limit the angular propagator $Q_{m}\left(  \theta_{b}%
,\theta_{a};S\right)  $ has the form%

\begin{equation}
Q_{m}\left(  \theta_{b},\theta_{a};S\right)  =\frac{1}{\sqrt{\sin\theta
_{b}\sin\theta_{a}}}\int D\theta\exp\frac{i}{\hbar}\int_{0}^{S}\left(
\frac{\mu}{2}\dot{\theta}^{2}-\frac{\hbar^{2}}{2\mu}\left(  \frac{\lambda
^{2}-\frac{1}{4}}{\sin^{2}\theta}+\frac{k^{2}-\frac{1}{4}}{\cos^{2}\theta
}\right)  \right)  ds \label{41}%
\end{equation}
which has some resemblance to the P\"{o}schl-Teller problem. $Q_{m}\left(
\theta_{b},\theta_{a};S\right)  $ is then integrable and the result is
\cite{PT1,PT2,PT3}%
\begin{equation}
Q_{m}\left(  \theta_{b},\theta_{a};S\right)  =\sum_{n_{\theta}=0}^{\infty
}e^{-\frac{i}{\hbar}\epsilon\left(  n_{\theta}\right)  S}~\Theta_{n_{\theta
},m}^{\ast}\left(  \theta_{a}\right)  ~\Theta_{n_{\theta},m}\left(  \theta
_{b}\right)  \label{42}%
\end{equation}
where $\epsilon\left(  n_{\theta}\right)  $ is given by
\begin{equation}
\epsilon\left(  n_{\theta}\right)  =\frac{\hbar^{2}}{2\mu}\left(  2n_{\theta
}+k+\lambda+1\right)  ^{2} \label{43}%
\end{equation}
and angular wave functions are%
\begin{equation}
\Theta_{n_{\theta},m}\left(  \theta\right)  =N_{n_{\theta},m}~\left(
\sin\theta\right)  ^{\lambda}\left(  \cos\theta\right)  ^{k+\frac{1}{2}%
}~P_{n_{\theta}}^{\left(  \lambda,k\right)  }\left(  \cos2\theta\right)
\label{44}%
\end{equation}
with the normalization constant%
\begin{equation}
N_{n_{\theta},m}=\sqrt{2\left(  2n_{\theta}+k+\lambda+1\right)  \frac
{n_{\theta}!\Gamma\left(  n_{\theta}+k+\lambda+1\right)  }{\Gamma\left(
n_{\theta}+k+1\right)  \Gamma\left(  n_{\theta}+\lambda+1\right)  }}
\label{45}%
\end{equation}
For $P_{E}\left(  r_{b},r_{a};S\right)  $ we can't do integration directly
because the kinetic term $\frac{\mu}{2r_{n}r_{n-1}}\frac{\left(  \Delta
r_{n}\right)  ^{2}}{\sigma_{n}}$ contains an inconvenient variable dependent
mass. It is the necessary to restore the original time $t.$ To this aim let us
proceed as follows. First we incorporate the result of $Q_{m}\left(
\theta_{b},\theta_{a};S\right)  $ in (\ref{36}) to get
\begin{equation}
\tilde{K}_{m}\left(  r_{b},\theta_{b};r_{a},\theta_{a};E\right)
=\sum_{n_{\theta}=0}^{\infty}\Theta_{n_{\theta},m}^{\ast}\left(  \theta
_{a}\right)  ~\Theta_{n_{\theta},m}\left(  \theta_{b}\right)  \mathcal{\tilde
{K}}_{n_{\theta},m}\left(  r_{b},r_{a};E\right)  ~ \label{46}%
\end{equation}
where $\mathcal{\tilde{K}}_{n_{\theta},m}\left(  r_{b},r_{a};E\right)  $ is
given by
\begin{equation}
\mathcal{\tilde{K}}_{n_{\theta},m}\left(  r_{b},r_{a};E\right)  =\int
_{0}^{+\infty}dS~~e^{-\frac{i}{\hbar}\epsilon\left(  n_{\theta}\right)
S}P_{E}\left(  r_{b},r_{a};S\right)  \label{47}%
\end{equation}
and has the following path integral representation%
\begin{align}
\mathcal{\tilde{K}}_{n_{\theta},m}\left(  r_{b},r_{a};E\right)   &
=\frac{r_{b}}{r_{a}}\int_{0}^{+\infty}dS~~\int\prod\limits_{n=1}^{N-1}%
dr_{n}\prod\limits_{n=1}^{N}\left(  \sqrt{\frac{\mu}{2i\pi\hbar\sigma_{n}%
r_{n}r_{n-1}}}\right) \nonumber\\
&  \exp\left\{  \frac{i}{\hbar}\sum_{n}\left[  \frac{\mu}{2}\frac{\left(
\Delta r_{n}\right)  ^{2}}{\sigma_{n}r_{n}r_{n-1}}-\left(  \frac{1}{2}%
m\omega^{2}r_{n}^{2}+\frac{\left(  \alpha-\beta\right)  \hbar^{2}}{2\mu
r_{n}^{2}}\right)  r_{n}r_{n-1}\sigma_{n}\right.  \right. \nonumber\\
&  \left.  \left.  +\left(  E+V_{0}\right)  r_{n}r_{n-1}\sigma_{n}-\left(
\frac{\hbar^{2}}{2\mu}\left(  2n_{\theta}+k+\lambda+1\right)  ^{2}-\frac
{\hbar^{2}}{8\mu}\right)  \sigma_{n}\right]  \right\}  . \label{48}%
\end{align}
Then we make the change $r_{n}r_{n-1}\sigma_{n}$ $\rightarrow$ $\varepsilon$
to obtain for $\mathcal{\tilde{K}}_{n_{\theta},m}\left(  r_{b},r_{a};E\right)
$ the following form
\begin{align}
\mathcal{\tilde{K}}_{n_{\theta},m}\left(  r_{b},r_{a};E\right)   &  =\frac
{1}{r_{b}r_{a}}\int_{0}^{+\infty}dT~~\int\prod\limits_{n=1}^{N-1}dr_{n}%
\prod\limits_{n=1}^{N}\left(  \sqrt{\frac{\mu}{2i\pi\hbar\varepsilon}}\right)
\nonumber\\
&  \exp\left\{  \frac{i}{\hbar}\sum_{n}\left(  \frac{\mu}{2}\frac{\left(
\Delta r_{n}\right)  ^{2}}{\varepsilon}-\left(  \frac{1}{2}m\omega^{2}%
r_{n}^{2}+\frac{\left(  \alpha-\beta\right)  \hbar^{2}}{2\mu r_{n}^{2}%
}\right)  \varepsilon\right.  \right. \nonumber\\
&  \left.  \left.  +\left(  E+V_{0}\right)  \varepsilon-\frac{\hbar^{2}}{2\mu
}\frac{\left(  2n_{\theta}+k+\lambda+1\right)  ^{2}-\frac{1}{4}}{r_{n}^{2}%
}\varepsilon\right)  \right\}  . \label{49}%
\end{align}
For $\mathcal{K}_{n_{\theta},m}\left(  r_{b},r_{a};T\right)  $ we have%
\begin{align}
\mathcal{K}_{n_{\theta},m}\left(  r_{b},r_{a};T\right)   &  =\frac{1}%
{r_{b}r_{a}}\int\prod\limits_{n=1}^{N-1}dr_{n}\prod\limits_{n=1}^{N}\left(
\sqrt{\frac{\mu}{2i\pi\hbar\varepsilon}}\right) \nonumber\\
&  \exp\left\{  \frac{i}{\hbar}\sum_{n}\left(  \frac{\mu}{2}\frac{\left(
\Delta r_{n}\right)  ^{2}}{\varepsilon}-\left(  -V_{0}+\frac{1}{2}m\omega
^{2}r_{n}^{2}+\frac{\left(  \alpha-\beta\right)  \hbar^{2}}{2\mu r_{n}^{2}%
}\right)  \varepsilon\right.  \right. \nonumber\\
&  \left.  \left.  -\frac{\hbar^{2}}{2\mu}\frac{\left(  2n_{\theta}%
+k+\lambda+1\right)  ^{2}-\frac{1}{4}}{r_{n}^{2}}\varepsilon\right)  \right\}
. \label{50}%
\end{align}
In the next section we show how to find solutions by integrating
$\mathcal{K}_{n_{\theta},m}\left(  r_{b},r_{a};T\right)  .$

\section{Integration of the radial propagator}

Having shown how to do separation of variables and integration over angular
ones let us do integration over radial variable to find the final solution to
our problem. Starting from the $\mathcal{K}_{n_{\theta},m}\left(  r_{b}%
,r_{a};T\right)  $ that can be written in the form
\begin{align}
\mathcal{K}_{n_{\theta}}\left(  r_{b},r_{a};T\right)   &  =\frac{1}{r_{b}%
r_{a}}\int\prod\limits_{n=1}^{N-1}dr_{n}\prod\limits_{n=1}^{N}\left(
\sqrt{\frac{\mu}{2i\pi\hbar\varepsilon}}\right) \nonumber\\
&  \exp\frac{i}{\hbar}\sum_{n}\left(  \frac{\mu}{2}\frac{\left(  \Delta
r_{n}\right)  ^{2}}{\varepsilon}-\left(  -V_{0}+\frac{1}{2}\mu\omega^{2}%
r_{n}^{2}+\frac{\hbar^{2}\tilde{\ell}\left(  \tilde{\ell}+1\right)  }{2\mu
r_{n}^{2}}\right)  \varepsilon\right)  , \label{51}%
\end{align}
with%
\begin{equation}
\tilde{\ell}=\sqrt{\left(  k+\lambda+2n_{\theta}+1\right)  ^{2}+\left(
\alpha-\beta\right)  }-\allowbreak\frac{1}{2} \label{52}%
\end{equation}
This path integral is the ordinary path integral of the radial oscillator, the
solution is known. We can find in \cite{P6} that $\mathcal{K}_{n_{\theta}%
}\left(  r_{b},r_{a};T\right)  $ takes the form%
\begin{align}
\mathcal{K}_{n_{\theta}}\left(  r_{b},r_{a};T\right)   &  =\frac{1}%
{\sqrt{r_{b}r_{a}}}e^{iV_{0}T}\frac{\mu\omega}{i\hbar\sin\left(  \omega
T\right)  }\nonumber\\
&  I_{\tilde{\ell}+\frac{1}{2}}\left(  \frac{\mu\omega r_{b}r_{a}}{\hbar
i\sin\left(  \omega T\right)  }\right)  \exp\left[  i\frac{\mu\omega}%
{2\hslash}\left(  r_{b}^{2}+r_{a}^{2}\right)  \cot\left(  \omega T\right)
\right]  . \label{53}%
\end{align}
Now to obtain the radial wave functions and the energetic levels we use the
Hille-Hardy formula \cite{grad}%
\begin{align}
&  \left.  \frac{s}{1-s^{2}}\exp\left[  -\frac{1}{2}\left(  X+Y\right)
\frac{1+s^{2}}{1-s^{2}}\right]  I_{\tilde{\ell}+\frac{1}{2}}\left(
\frac{2\sqrt{XY}s}{1-s^{2}}\right)  =\right. \nonumber\\
&  \sum_{n=0}^{\infty}\frac{s^{2n+\tilde{\ell}+\frac{3}{2}}n!e^{-\frac{1}%
{2}\left(  X+Y\right)  }}{\Gamma\left(  n+\alpha+1\right)  }\left(  \sqrt
{XY}\right)  ^{\tilde{\ell}+\frac{1}{2}}L_{n}^{\tilde{\ell}+\frac{1}{2}%
}\left(  X\right)  L_{n}^{\tilde{\ell}+\frac{1}{2}}\left(  Y\right)  ,
\label{54}%
\end{align}
and we take $s=e^{-i\omega T};~\ \ \ X=\frac{\mu\omega}{\hslash}r_{b}%
^{2};~\ \ \ \ Y=\frac{\mu\omega}{\hslash}r_{a}^{2},$ to get the spectral
decomposition of $\mathcal{K}_{n_{\theta},m}\left(  r_{b},r_{a};T\right)  $%
\begin{align}
\mathcal{K}_{n_{\theta},m}\left(  r_{b},r_{a};T\right)   &  =2\left(
\frac{\mu\omega}{\hbar}\right)  ^{\frac{3}{2}}\sum_{n=0}^{\infty}e^{-i\left[
\omega\left(  2n+\tilde{\ell}+\frac{3}{2}\right)  -V_{0}\right]  T}\frac
{n!}{\Gamma\left(  n+\tilde{\ell}+\frac{3}{2}\right)  }\times\nonumber\\
&  \exp\left(  -\frac{\mu\omega}{2\hslash}r_{b}^{2}-\frac{\mu\omega}{2\hslash
}r_{a}^{2}\right)  \left(  \frac{\mu\omega}{\hslash}r_{b}r_{a}\right)
^{\tilde{\ell}}L_{n}^{\tilde{\ell}+\frac{1}{2}}\left(  \frac{\mu\omega
}{\hslash}r_{b}^{2}\right)  L_{n}^{\tilde{\ell}+\frac{1}{2}}\left(  \frac
{\mu\omega}{\hslash}r_{a}^{2}\right)  \label{55}%
\end{align}
or in closed form%
\begin{equation}
\mathcal{K}_{n_{\theta},m}\left(  r_{b},r_{a};T\right)  =\sum_{n=0}^{\infty
}e^{-\frac{i}{\hbar}E_{n,n_{\theta},m}T}\times R_{n,n_{\theta},m}^{\ast
}\left(  r_{a}\right)  R_{n,n_{\theta},m}\left(  r_{b}\right)  \label{56}%
\end{equation}
where the energy spectrum is given by%
\begin{equation}
E_{n,n_{\theta},m}=\left(  2n+\tilde{\ell}+\frac{3}{2}\right)  \hbar
\omega-V_{0} \label{57}%
\end{equation}
and the radial wave function reads%
\begin{equation}
R_{n,n_{\theta},m}\left(  r\right)  =\sqrt{2\left(  \frac{\mu\omega}{\hbar
}\right)  ^{\frac{3}{2}}\frac{n!}{\Gamma\left(  n+\tilde{\ell}+\frac{3}%
{2}\right)  }}\exp\left(  -\frac{\mu\omega}{2\hslash}r^{2}\right)  \left(
\sqrt{\frac{\mu\omega}{\hslash}}r\right)  ^{\tilde{\ell}}~\ L_{n}^{\tilde
{\ell}+\frac{1}{2}}\left(  \frac{\mu\omega}{\hslash}r^{2}\right)  \label{58}%
\end{equation}
Finally the relative propagator can be expressed in the following spectral
decomposition%
\begin{equation}
K\left(  \vec{r}_{b},\vec{r}_{a};T\right)  =\sum_{n=0}^{\infty}\sum
_{n_{\theta}=0}^{\infty}\sum_{m=-\infty}^{+\infty}e^{-\frac{i}{\hbar
}E_{n,n_{\theta},m}T}\psi_{n,n_{\theta},m}^{\ast}\left(  \vec{r}_{a}\right)
\psi_{n,n_{\theta},m}\left(  \vec{r}_{b}\right)  , \label{59}%
\end{equation}
where%
\begin{align}
\psi_{nlm}\left(  \vec{r}\right)   &  =N_{n,n_{\theta},m}\exp\left(
-\frac{\mu\omega}{2\hslash}r^{2}\right)  \left(  \sqrt{\frac{\mu\omega
}{\hslash}}r\right)  ^{\tilde{\ell}}~\ L_{n}^{\tilde{\ell}+\frac{1}{2}}\left(
\frac{\mu\omega}{\hslash}r^{2}\right) \nonumber\\
&  \times\left(  \sin\theta\right)  ^{\lambda}\left(  \cos\theta\right)
^{k+\frac{1}{2}}~P_{n_{\theta}}^{\left(  \lambda,k\right)  }\left(
\cos2\theta\right)  e^{im\varphi}, \label{60}%
\end{align}
with the normalization constant%
\begin{equation}
N_{n,n_{\theta},m}=\sqrt{\frac{2}{\pi}\left(  \frac{\mu\omega}{\hbar}\right)
^{\frac{3}{2}}\left(  2n_{\theta}+k+\lambda+1\right)  \frac{n!n_{\theta
}!\Gamma\left(  n_{\theta}+k+\lambda+1\right)  }{\Gamma\left(  n_{\theta
}+k+1\right)  \Gamma\left(  n_{\theta}+\lambda+1\right)  \Gamma\left(
n+\tilde{\ell}+\frac{3}{2}\right)  }}. \label{61}%
\end{equation}
Let us remark that when we consider the particular cases studied in
\cite{NC1,NC2,NC3}, we see that our results coincide with these results.

\section{Conclusion}

In this paper we have given a straightforward method to solve the problem of
noncentral anharmonic oscillator in three dimensions. In the first stage we
have expressed the relative propagator by means of path integrals in spherical
coordinates. Then by making an adequate change of time we were able to
separate the angular motion from the radial one. The angular part path
integration is reduced to the well-known P\"{o}schl-Teller problem and the
resulting radial path integral is written in the form of three dimensional
isotropic harmonic oscillator. Then we have exactly calculated the relative
propagator and we have extracted the bound states energies and the
corresponding wave functions.

We remark also that the presented method is useful for all potentials of the
form%
\begin{equation}
V\left(  r,\theta\right)  =v\left(  r\right)  +~\frac{u\left(  \theta\right)
}{r^{2}} \label{62}%
\end{equation}
and there is no need to use other systems of coordinates.

Through the formulation given above and the obtained energies and wave
functions we conclude that the path integral formulation is a powerful method
to study quantum dynamics of particles in nonrelativistic theory.


\begin{thebibliography}{99}                                                                                               %


\bibitem {P1}Feynman R. P. and Hibbs A. R., Quantum Mechanics and Path
Integrals (Mc Graw Hill, New York, 1965).

\bibitem {P2}Schulman L. S., Techniques and Applications of Path Integration
(John Wiley, New York, 1981)

\bibitem {P3}Dittrich W. and Reuter M., Classical and Quantum Dynamics: From
Classical Paths to Path Integrals (Springer, Berlin, 2001)

\bibitem {P4}Khandekar D.C. and Lawande S.V., Phys. Rep. \textbf{137,} 115 (1986)

\bibitem {P5}Khandekar D. C., Lawande S. V. and Bhagwat K. V., Path Integral
methods and their applications (World scientific, singapore 1993)

\bibitem {P6}Grosch C. and Steiner F., Handbook of Feynman Path Integrals,
Springer Tracts in Modern Physics 145 ( Springer, Berlin, Heidelberg 1998)

\bibitem {Duru}Duru H. and Kleinert H., Fortschr. Phys. \textbf{30,} 401 (1982).

\bibitem {P7}Kleinert H., Path Integral in quantum mechanics, statistics and
polymer physics (World Scientific, Singapore 1990).

\bibitem {P8}Ashok Das, Field Theory: A Path Integral Approach (World
Scientific singapore 2006)

\bibitem {P9}Ulrich Mosel, Path Integrals in Field Theory: An Introduction
(Springer,~Berlin 2003)

\bibitem {P10}Zinn-Justin, J., Quantum Field Theory and Critical Phenomena,
fourth edition (Oxford University Press, New York, 2002).

\bibitem {P11}Herbert W. Hamber, Quantum Gravitation; The Feynman Path
Integral Approach (Springer-Verlag, Berlin, 2009)

\bibitem {P12}Chitre D. M., and Hartle J. B., Phys. Rev D. \textbf{16}, 251 (1977).

\bibitem {P13}Hawking S. W., and Hartle J. B., Phys. Rev. D \textbf{13,} 2188 (1976).

\bibitem {Chaichian1}Chaichian M. and Demichev A., Path integrals in physics.
\textbf{Vol.1}: Stochastic processes and quantum mechanics (IOP Publisher,
Bristol UK 2001).

\bibitem {Chaichian2}Chaichian M. and Demichev A., Path integrals in physics.
\textbf{Vol.2}: Quantum field theory, statistical physics and other modern
applications, (IOP Publisher, Bristol UK 2001).

\bibitem {setare}Setare M. R. and Haidari S., Int. J. Theor. Phys.
\textbf{48,} 3249 (2009).

\bibitem {Grosch}Grosch C., Fortsch.Phys. \textbf{43,} 453 (1995).

\bibitem {NC1}Aktas M., Int. J. Theor. Phys. \textbf{48,} 2154 (2009).

\bibitem {NC2}Dong S-H., Sun G-H., and Lozada-Cassou M., Phys. Lett. A
\textbf{340, }94 (2005).

\bibitem {NC3}Berkdemir C., J Math Chem \textbf{46,} 139 (2009).

\bibitem {NC4}Zhang Xue-Ao, Chen Ke and Duan Zheng-Lu, Chin. Phys. \textbf{14,
}42 (2005)

\bibitem {NC5}Lu Fa-Lin, Chen Chang-Yuan and Sun Dong-Sheng, Chin. Phys.
\textbf{14,} 464 (2005)

\bibitem {NC6}C. Berkdemir and Yan-Fu Cheng, Phys. Scr. \textbf{79,} 035003 (2009)

\bibitem {NC7}Gao-FengWei, Chao-Yun Long, Zhi He, Shui-Jie Qin and Jing Zhao,
Phys. Scr. \textbf{76,} 442 (2007)

\bibitem {NC8}Min-Cang Zhang, Guo-Qing Huang-Fu and Bo An, Phys. Scr.
\textbf{80,} 065018 (2009)

\bibitem {mclaughlin}McLaughlin D. and Schulman L.S., J. Math. Phys.
\textbf{12,} 2520 (1971).

\bibitem {chetouani}Chetouani L., Guechi L. and Hammann T. F., J. Math. Phys.
\textbf{30,} 655 (1989)

\bibitem {PT1}Duru I. H., Phys. Rev. D \textbf{30,} 2121 (1984)

\bibitem {PT2}Inomata A. and Wilson R., Path Integral Realization of a
Dynamical Group; Lecture Notes in Physics 261 (Springer, Berlin-Heidelberg, 1985)

\bibitem {PT3}B\"{o}hm M. and Junker G., J. Math. Phys. \textbf{28,} 1978 (1987)

\bibitem {grad}Gradshteyn I. S. and Ryzhik I. M., Table of Integrals, Series,
and Products (Academic Press, New York 1979)
\end{thebibliography}
\end{document}